\documentclass[]{aa}
\usepackage{graphicx}
\usepackage{natbib}
\usepackage{ulem}
\begin{document}

\title{Multiline Zeeman Signatures through line addition} 

 \author{M. Semel\inst{1}, J.C. Ram\'{i}rez V\'elez\inst{1}, M.J. Mart\'{i}nez Gonz\'alez\inst{2,3}, A. Asensio Ramos\inst{3}, 
M.J.~Stift\inst{2,4}, A. L\'opez Ariste\inst{5} \& F. Leone\inst{6}}

\institute{
LESIA, Observatoire de Paris Meudon. 92195 Meudon,
France. \email{Meir.Semel@obspm.fr, Julio.Ramirez@obspm.fr} 
\and LERMA, Observatoire de Paris Meudon. 92195 Meudon, France. \email{martin.stift@univie.ac.at, marian@iac.es} 
\and Instituto de Astrof\' isica de Canarias, V\' ia L\'actea s/n, E-38205, La Laguna, Spain
\and Institute for Astronomy, Univ. of Vienna, T{\"u}rkenschanzstrasse 17,
1180 Vienna, Austria.
\and THEMIS, CNRS UPS 853, c/v\'{\i}a L\'actea s/n. 38200. La  Laguna,
Tenerife, Spain; \email{arturo@themis.iac.es}
\and INAF, Osservatorio Astrofisico di Catania, Via S. Sofia 
n 78,95123 Catania, Italy
}

\offprints{meir.semel@obspm.fr}

\titlerunning{Multiline Zeeman Signatures}
\authorrunning{Meir Semel et al.\ \ }
 \date {Received date , accepted date}

\begin{abstract}
{In order to get a significant Zeeman signature in the polarised
spectra of a magnetic star, we usually 'add' the contributions of numerous
spectral lines; the ultimate  goal is to recover the spectropolarimetric
prints of the  magnetic field in these line additions.}  
{Here we want to clarify the meaning of these techniques of line
addition; in particular, we try to interpret the meaning of the 
'pseudo-line' formed during this process and to find out why and
how its Zeeman signature is still meaningful.}
{We create a synthetic case of line addition and apply well tested
standard solar methods routinely used in the research on magnetism 
in our nearest star.}
{The results are convincing and the Zeeman signatures well detected;
Solar methods are found to be quite efficient also for stellar 
observations. We statistically compare line addition with least-squares deconvolution
and demonstrate that they both give very similar results as a consequence of
the special statistical properties of the weights.}
{The Zeeman signatures are unequivocally detected in this multiline 
approach. We may anticipate the outcome that magnetic field detection is reliable 
well beyond the weak-field approximation. Linear polarisation in the spectra of 
solar type stars can be detected when the spectral resolution is 
sufficiently high.}
\end{abstract}

\keywords{}

\maketitle

\section{On the Multiline Zeeman Signature and Zeeman Doppler Imaging concepts}

In the 1980s it became clear that cool stars, mainly rapid rotators,
exhibit solar type activity. Rapid rotation is one of the essential
ingredients to the {\it stellar dynamo}. While there is not yet any
completely satisfactory theory or model of the solar or the stellar
dynamo, it is quite clear what drives such a process, viz. convection,
rapid rotation and differential rotation. Tentatives to detect the
Zeeman effect in these active stars however failed for two main reasons:
\begin{itemize}
\item The magnetic configuration of the stellar magnetic fields can
be quite complex. The simultaneous appearance of opposite magnetic
polarities may lead to cancellation of the respective contributions
to the integrated polarisation signal in the stellar spectrum.
\item The broadening of the spectral lines due to even moderate stellar
rotation rates can become a serious handicap when measuring the
magnetic field of the star.
\end{itemize}
Consideration of these two aspects was at the object of Semel (1989)
where the term Zeeman Doppler Imaging (ZDI) stood for the detection of a Zeeman
signature thanks to the Doppler effect. The latter can help to
disentangle the contributions from opposite magnetic polarities and
their respective opposite polarisation which otherwise may cancel.
Thus, the combination of Zeeman and Doppler effects is one of the
reasons why the detection of the Zeeman effect in fast rotating active
solar type stars becomes possible. Nowadays, however, the term ZDI is
often understood as magnetic mapping of the stellar surface, based on
the inversion of Stokes profiles (frequently only $I$ and $V$) sampled
over a full period of rotation.

In order to overcome possible misunderstandings, we therefore propose the 
term {\em Multiline Zeeman Signature} (henceforth MZS) to denote a method for just the detection of a mean
Zeeman signature using numerous spectral lines.
We shall not attribute any direct physical meaning to the MZS.
We simply require the application of an operator {\sf O} (or a detector 
{\sf O}) to a polarised spectrum, creating a particular MZS$_\mathsf{O}$. 

Nowadays, there are several techniques that take into account many 
spectral lines to obtain some particular MZSs that trace 
stellar magnetic fields. One of the most elementary methods is the so called \emph{line addition} 
technique which consists of simply adding up many observed spectral lines (Semel 1989, Semel \& Li 1996). 
Later on, Donati et al. (1997) introduced the Least Squares Deconvolution (LSD) method. 
More recently, methods based on Principal Component Analysis (PCA) 
have been introduced by Semel et al. (2006) and Mart\' inez Gonz\'alez et. al (2008). In this paper we 
concentrate on the proper definition of the line addition technique and its application to any of the 
Stokes parameters. We present the advantages of using this technique as compared to LSD and 
we describe the correct way to infer physical properties of its MZS.

\section{The MZS through line addition technique as compared to the LSD profile}
\subsection{Line Addition}
The first successful MZS tentatives consisted in adding up 
the circular polarisation of selected spectral lines. In the 
first efforts, Semel (1989) recommended the de-blending of spectral lines prior to
addition, following a relatively tough procedure. However, Semel \& Li (1996) showed 
that simple coherent addition 
of spectral lines gave satisfactory results. This is due to the fact that the 
the position of the blends along the spectral lines is not systematic and they 
add incoherently. Each spectral line may appear several times, 
once in the coherent addition, and then each time when the line is 
``blending'' in the field of another nearby line. However, in the
latter case it appears in a non-coherent way and always in another 
position (in wavelength). When the line is blending, its contribution
is ``arbitrarily'' situated and has only a little effect. In other words, 
the blending lines only contribute to the noise, perhaps modifying its statistical
distribution. From this point, noise is reduced when the number of spectral lines increases.

In the above mentioned works, some 200 spectral lines were added in order to 
yield significant detections of Zeeman signatures. This straightforward method of line 
addition was later improved by Semel (1995) in which a list of spectral lines of interest 
was created. These spectral lines were labelled by wavelength, $\lambda_i$,
equivalent width, $w_i$, and effective Land{\'e} factor, $\bar g_i$. In the
next step, each spectral line was represented by a Dirac function
$w_i \cdot \bar g_i \cdot \delta(\lambda - \lambda_i)$ and transformed, as
explained in Semel (1989), to the Doppler coordinate $X$, with
$dX = c \, d\lambda/\lambda$, being $c$ the velocity of light. Using
$\omega_i^\mathrm{LA} = w_i \bar g_i$, the
convolution of the observed circular polarisation spectrum $V(X)$
and the line list results in a MZS given by:
\begin{equation}
Z_\mathrm{LA}(x_j)= \frac{\sum_i\,\omega_i^\mathrm{LA} \,V(x_j-X_i)}{\sum_i\,\omega_i^\mathrm{LA}},
\label{eq:LA}
\end{equation}
where $X_i$ is the Doppler coordinate of the centre of each spectral line and 
$x_j$ is the Doppler coordinate axis. The symbol $Z_\mathrm{LA}(x_j)$ stands
for the MZS due to line addition.
As a rule, the convolution by means of a Fourier transform is found
to be quite efficient. The line addition technique has also been applied
with unit weights, i.e., $\omega_i^\mathrm{LA}=1$, with very similar results
to the more refined version presented above:
\begin{equation}
Z_\mathrm{SLA}(x_j) = \frac{1}{N} \sum_i V(x_j-X_i),
\label{eq:simplified_LA}
\end{equation}
where SLA stands for simplified line addition.

Although initially it was only used for the circular state of polarisation,
since the MZS resulting from line addition is model-independent, it can be 
applied either to Stokes $I$, $Q$, $U$ or $V$. The reason is that no assumption 
have been made in order to compute the MZS, being just the brute force line addition. 
Therefore, the line addition is a very powerful and robust method to detect Zeeman signatures in stellar 
polarised spectra where the signal-to-noise ratio per individual spectral line is very 
poor. 

\subsection{Least-squares Deconvolution}
The LSD technique was introduced for the detection of Zeeman signatures in 
stellar polarised spectra by Donati et al. (1997). They postulated
that, in a given observed Stokes spectrum, the shape of the local Zeeman signature in
circular polarisation is identical for all spectral lines. This means that individual
Stokes $V$ profiles are assumed to correspond to a common basic
Zeeman signature that is modulated by a proportionality factor. This coefficient 
is then fixed in the frame of the weak field approximation. In other words, they assume that 
the magnetic field is weak enough so that the Zeeman splitting is negligible as compared 
to the Doppler width of the spectral line. As a consequence, the radiative transfer
equation can be solved analytically and it can be demonstrated that the Stokes $V$ profile
is proportional to the derivative of the intensity profile (Sears 1913; Landi degl'Innocenti \& Landolfi 2004). Following this approximation, 
Donati et al. (1997) found that the proportionality factor, $\omega_i^\mathrm{LSD}=\lambda_i \, d_i \, \bar g_i$, i.e., 
just the product of the central 
wavelength of the transition, $\lambda_i$, the central depth of the line, $d_i$, and the effective Land\'e factor, $\bar g_i$.
Donati et al. (1997) also normalised the weights to a mean value of 500 nm.
Based on all these assumptions, a particular MZS is extracted by means of
a least squares method from the observed spectrum. 

We note in passing that $\omega_i^\mathrm{LSD}$ and $\omega_i^\mathrm{LA}$, although apparently
different, have a very similar behaviour. If the spectral line is assumed to be of Gaussian shape
whose width is dominated by thermal effects, it is possible to show that 
$\omega_i^\mathrm{LA} \propto \omega_i^\mathrm{LSD}$ in general.

In more detail, the MZS obtained through the least-squares deconvolution method of Donati et al. (1997) is
calculated solving the following weighted linear regression problem: given $N$ lines with Stokes $V$
profiles $V(x_j-X_i)$ sampled at the $N_x$ Doppler coordinates $x_j$, find the profile 
$Z(x_j)$ that better approximates the observations in the least-squares sense 
given the fixed values of $\{\omega_i^\mathrm{LSD}\}$. In other words, it can be seen as the problem of fitting $N_x$
independent straight lines whose slopes are given by the values of $Z(x_j)$ and the 
abscissae are the values of the weights $\{\omega_i^\mathrm{LSD}\}$. The solution is obtained, under the
case of Gaussian noise, by minimising the following merit functions:
\begin{equation}
\chi^2_j = \sum_i W_i \left[V(x_j-X_i)-\omega_i^\mathrm{LSD} Z(x_j) \right]^2,
\end{equation}
where the weights are usually chosen to be $W_i=1/\sigma(x_j-X_i)^2$, the inverse variance of the expected noise at each
velocity point. The solution to the previous problem is:
\begin{equation}
Z_\mathrm{LSD}(x_j) = \frac{\sum_i W_i \omega_i^\mathrm{LSD} V(x_j-X_i)}{\sum_i W_i ({\omega_i^\mathrm{LSD}})^2}.
\label{eq:LSD}
\end{equation}

\begin{figure*}[!t]
\center
\includegraphics[width=0.49\textwidth]{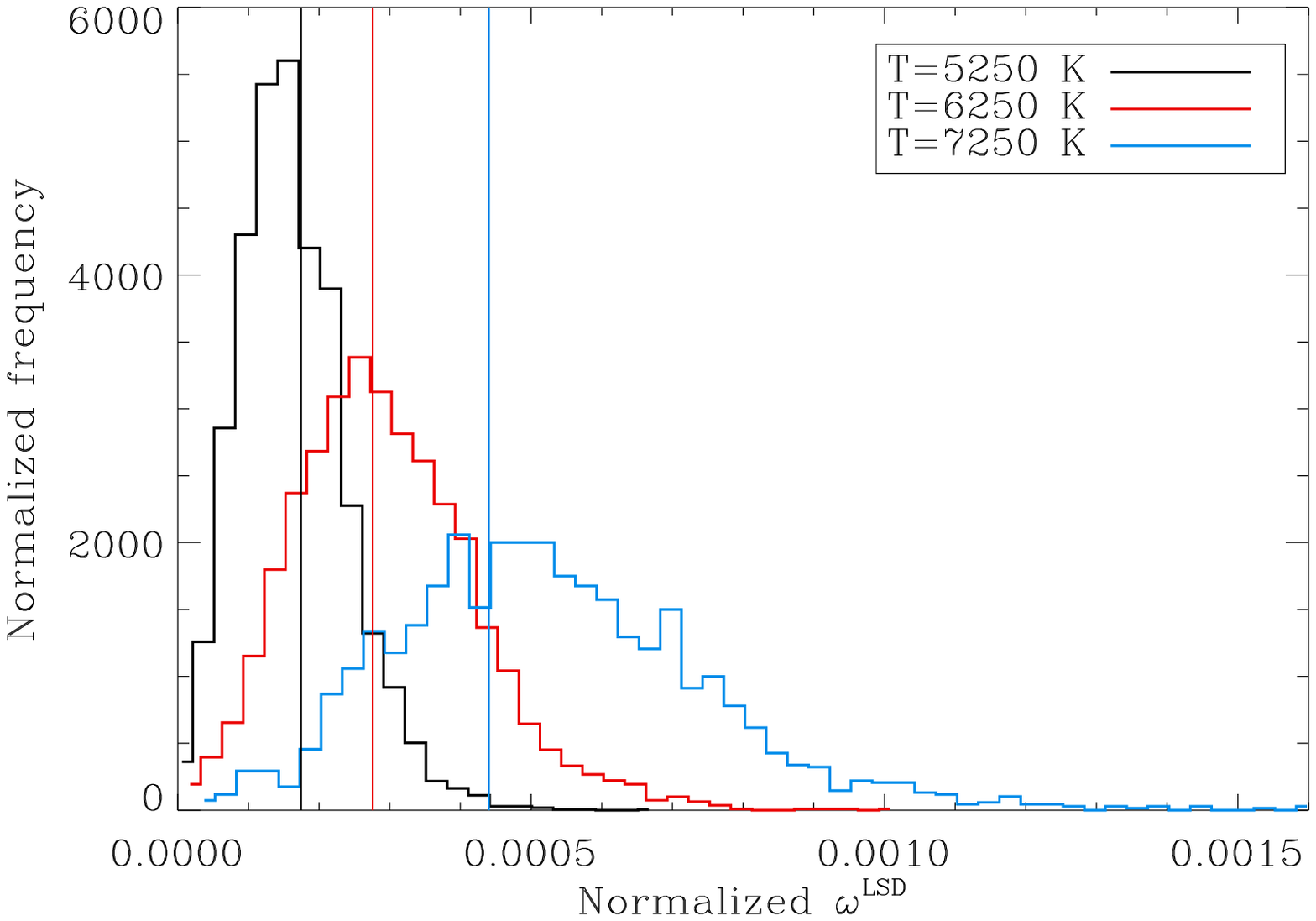}
\includegraphics[width=0.49\textwidth]{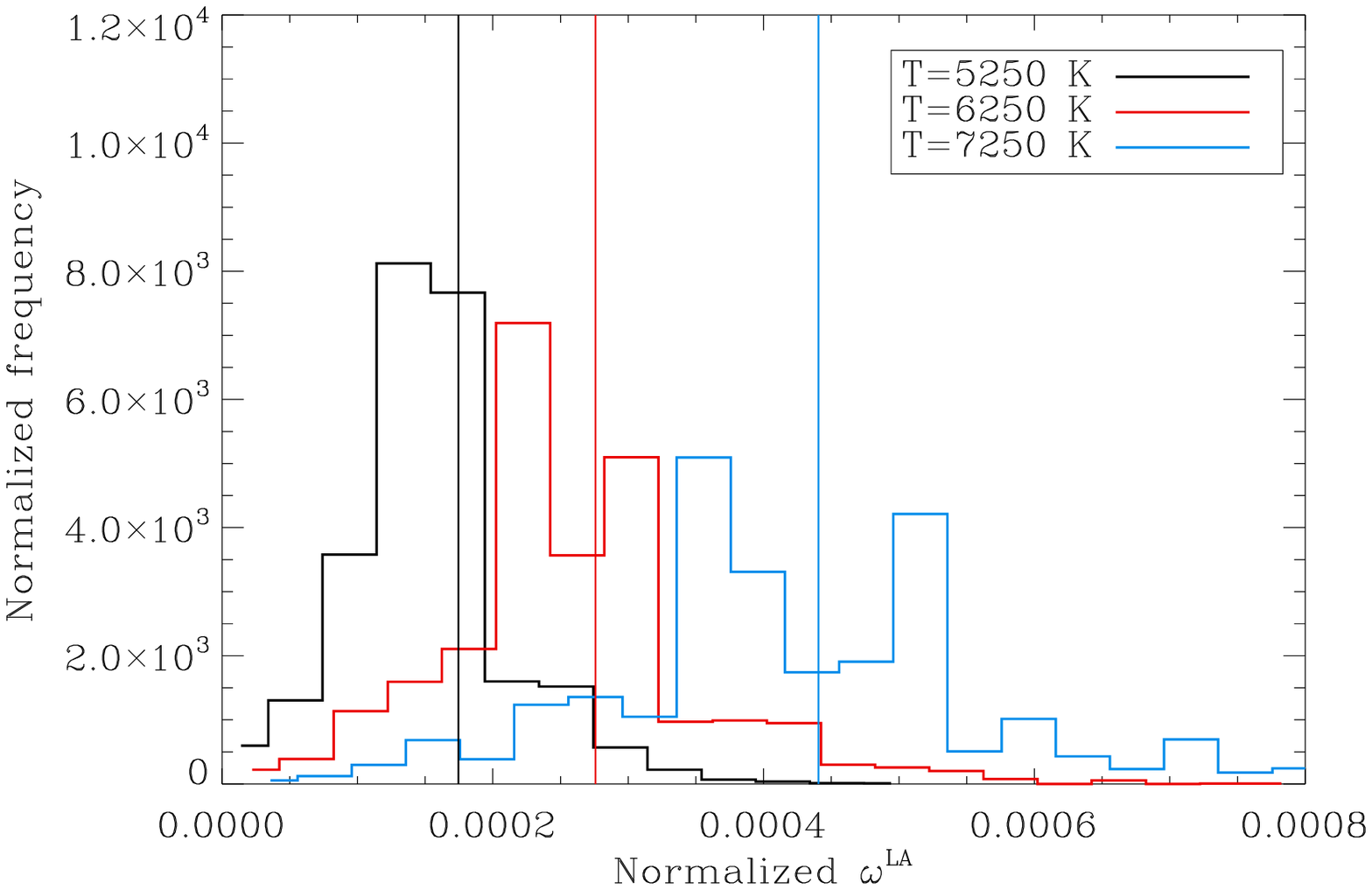}
\caption{The left panel shows the probability distribution of the normalised LSD weights, 
$\omega_i^\mathrm{LSD} / \sum_i ({\omega_i^\mathrm{LSD}})^2$. The vertical lines indicate the position
of $1/N$ for each temperature, where $N$ is the number of lines included in the histogram.
The right panel presents the probability distribution function of the normalised simplified 
line addition weights, $\bar g_i/ \sum_i \bar g_i$.}
\label{fig:weights}
\end{figure*}

In combination with
maximum entropy codes (Brown et al. 1991, Donati \& Brown 1997), LSD-based 
Stokes $I$ and $V$ profiles have been used for the production of
magnetic surface maps of quite a large number of stars with great success.
 
The main disadvantage of the LSD method relies on the assumptions made to obtain the MZS. 
It is not true in general that, locally, all spectral lines have a common shape (see for instance Kurucz 1993). 
Moreover, the weak field approximation used to obtain the proportionality coefficients states that 
Stokes $Q$ and $U$ are zero to first order. This would make this technique not applicable to recover the MZS of the linear 
states of polarisation. 


\subsection{Statistical properties of weights}
A fundamental point to clarify is why LSD works even though it is based on assumptions
that may be difficult to admit. In order to shed some light on this issue, it turns out
essential to analyse the statistical properties of the weights. The left panel 
of Fig. \ref{fig:weights} presents the probability distribution function of the normalised weights
$\omega_i^\mathrm{LSD} / \sum_i ({\omega_i^\mathrm{LSD}})^2$ for many lines in several line-lists obtained for different
stellar temperatures. We have verified that the distribution is quite insensitive to 
the value of the surface gravity and the only strong dependency is seen with respect
to the effective temperature. The histograms have been calculated for lines above 300 nm 
and whose calculated line depth is larger than 0.4, which is the typical selection
criterium used in the standard application of LSD. The vertical lines in Fig. \ref{fig:weights}
are the value of the inverse of the number of lines included in the histograms, in other
words, the value of the weights in the simplest version of line addition given by
Eq. (\ref{eq:simplified_LA}). Therefore, 
the LSD weights can be considered, to a good approximation, to be positive random 
variables whose distribution is close
to log-normal (which can be not badly represented by a Gaussian distribution), centred around 
$1/N$ and with a certain dispersion. The log-normal shape arises 
because the product of several independent random quantities converge towards
such a distribution, similar to the convergence towards a Gaussian distribution when
adding many independent quantities. In our case, the weights are the product of three quantities 
that do not present a clear dependency (surely not the effective Land\'e factor and the 
rest of quantities).

The right panel of Fig. \ref{fig:weights} presents a similar statistical analysis but
for line addition. In our study we use standard line-lists applied to LSD in which the
equivalent width is not present. For this reason, we plot the weights for the
simplified case in which the weights are just the effective Land\'e factors of all
included lines, so that we plot the histogram of $\bar g_i/ \sum_i \bar g_i$. The
vertical lines again indicate the value of $1/N$ for each temperature. The weights
can again be represented approximately as Gaussian random variables centred around
$1/N$ and with a certain dispersion.

\begin{figure*}[!t]
\center
~~~~500 G~~~~~~~~~~~~~~~~~~~~~~~~~~~~~~~~~~~~~~~~~~~~~1500 G~~~~~~~~~~~~~~~~~~~~~~~~~~~~~~~~~~~~~~~~~~~2500 G\\
\includegraphics[width=0.31\textwidth]{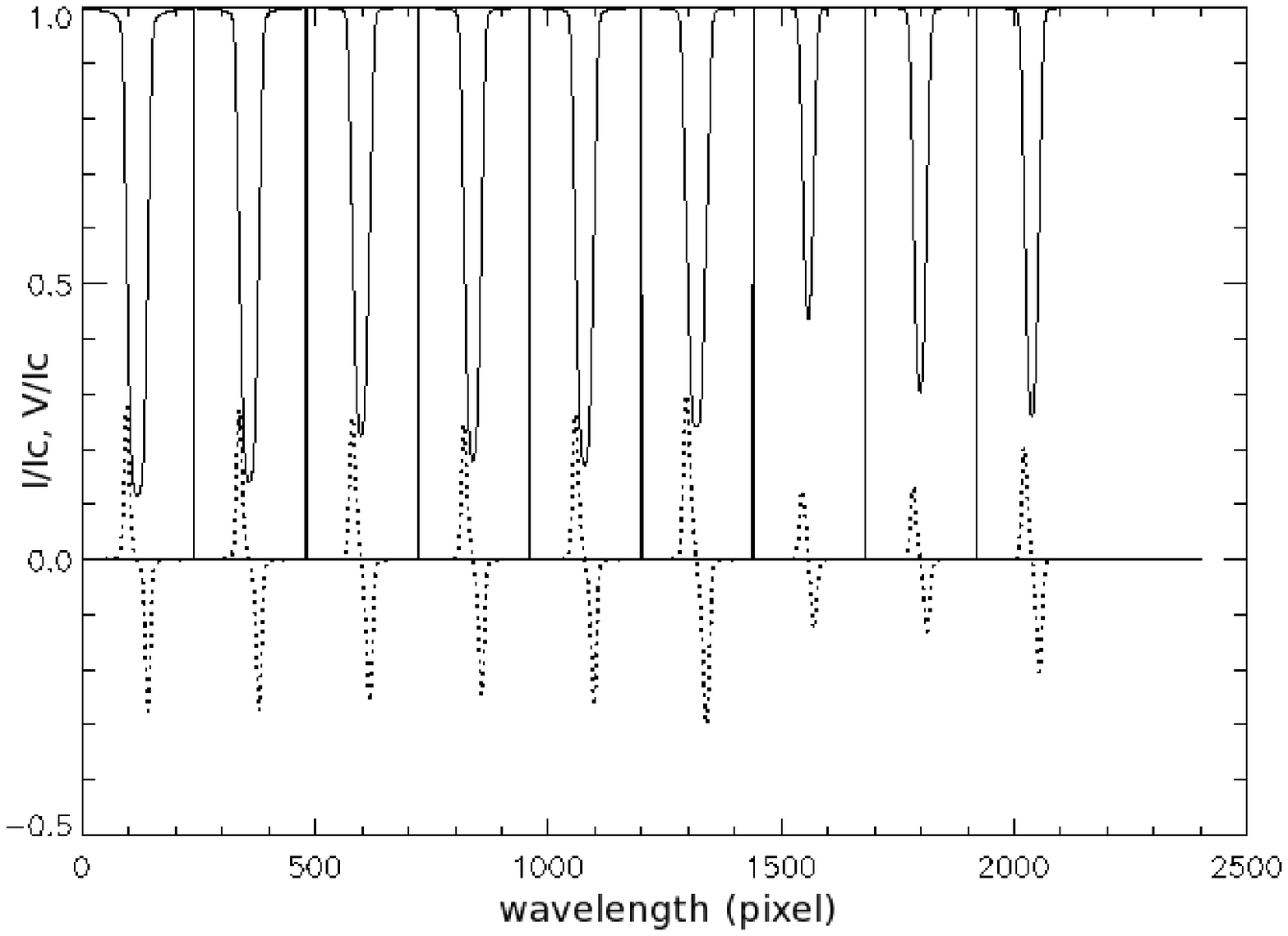}
\includegraphics[width=0.31\textwidth]{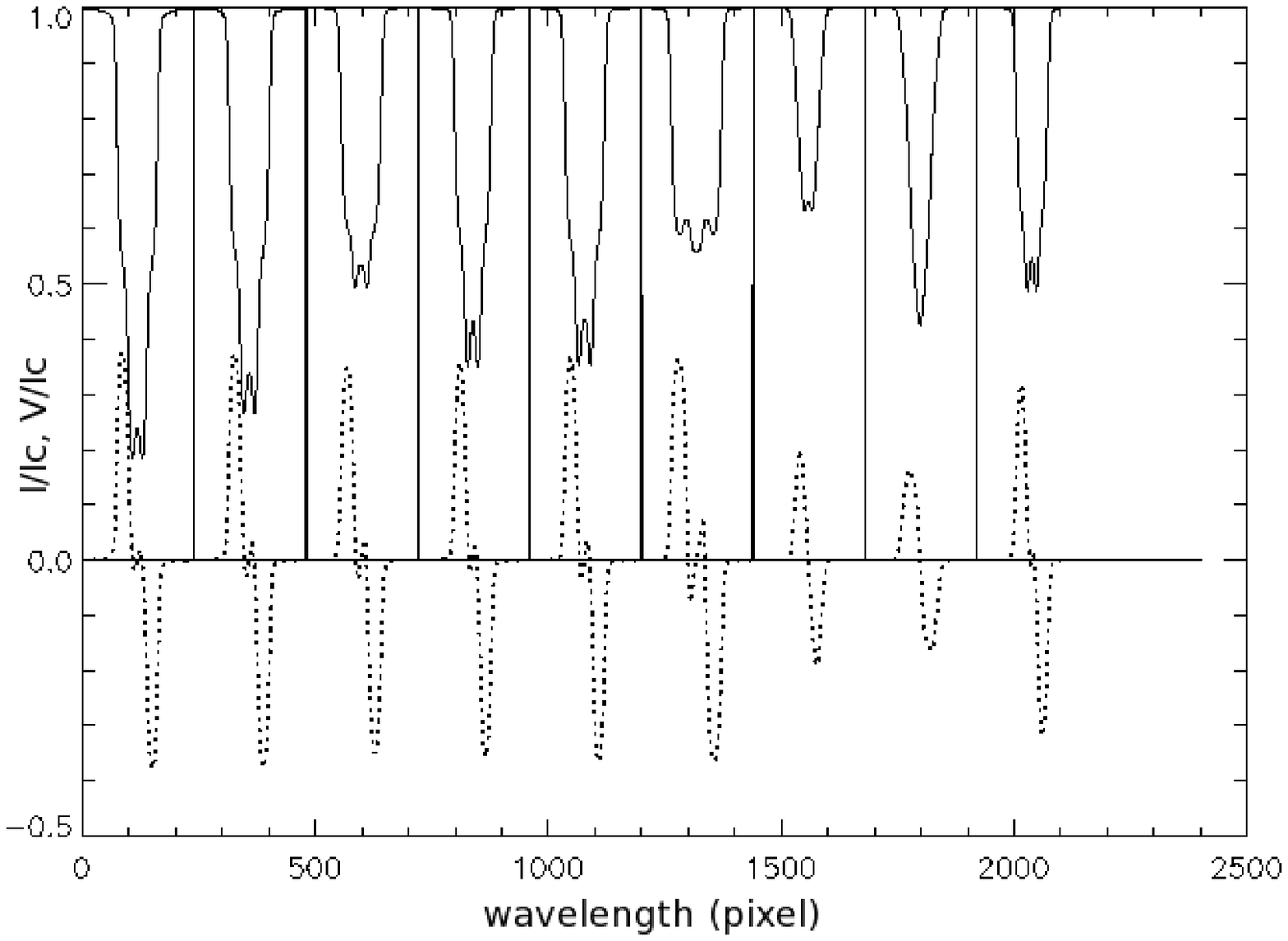}
\includegraphics[width=0.31\textwidth]{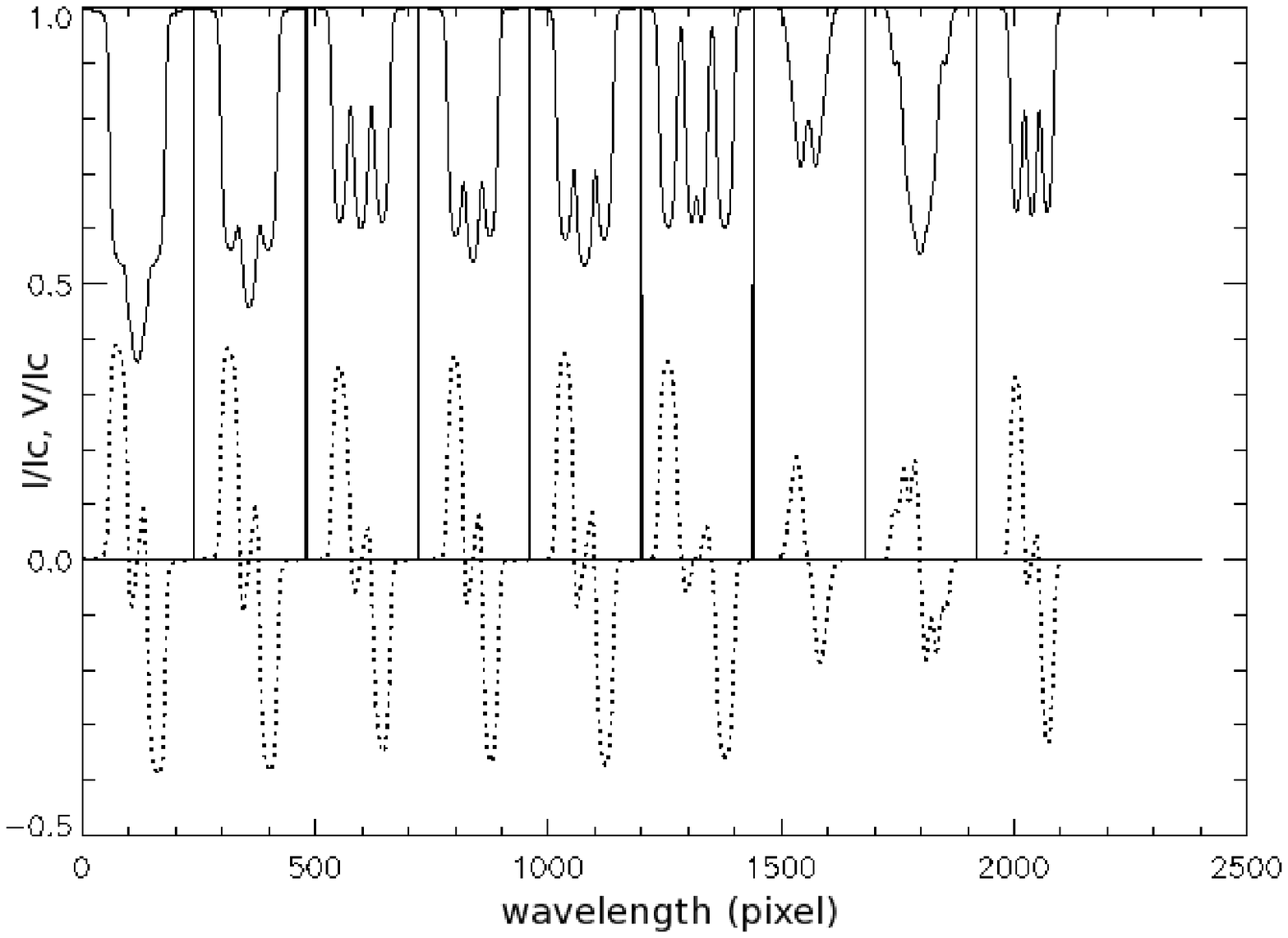}\\
\includegraphics[width=0.31\textwidth]{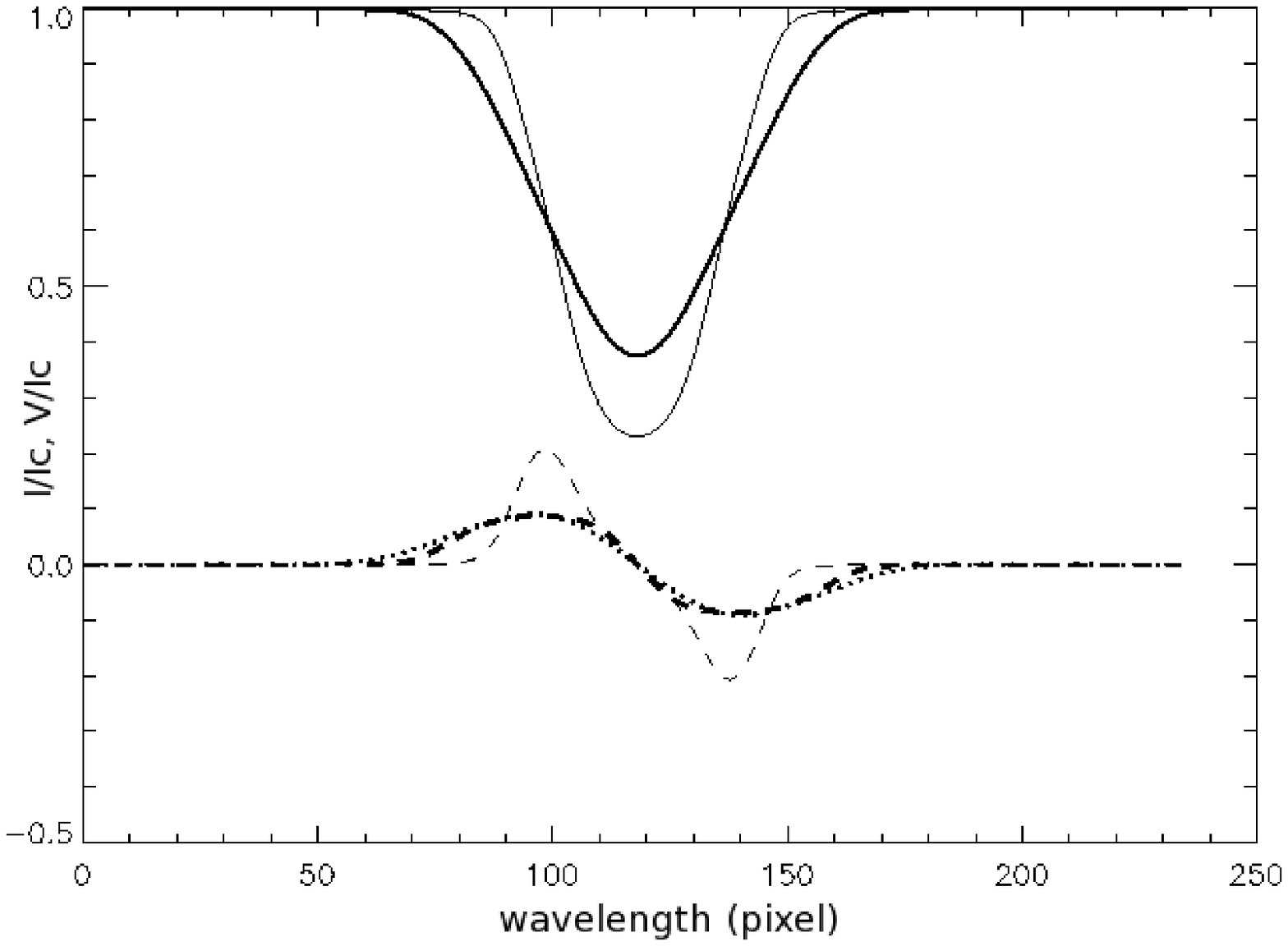}
\includegraphics[width=0.31\textwidth]{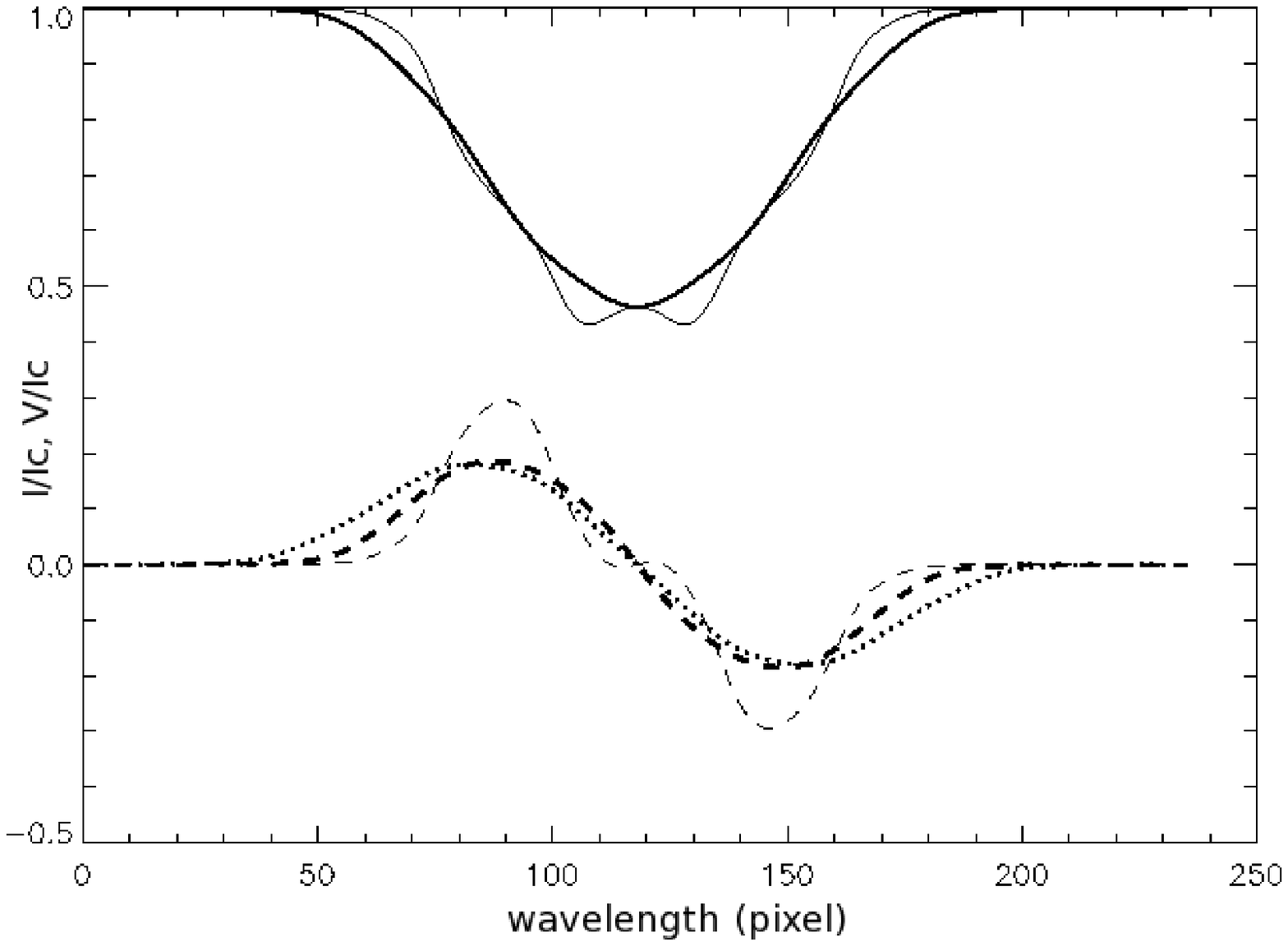}
\includegraphics[width=0.31\textwidth]{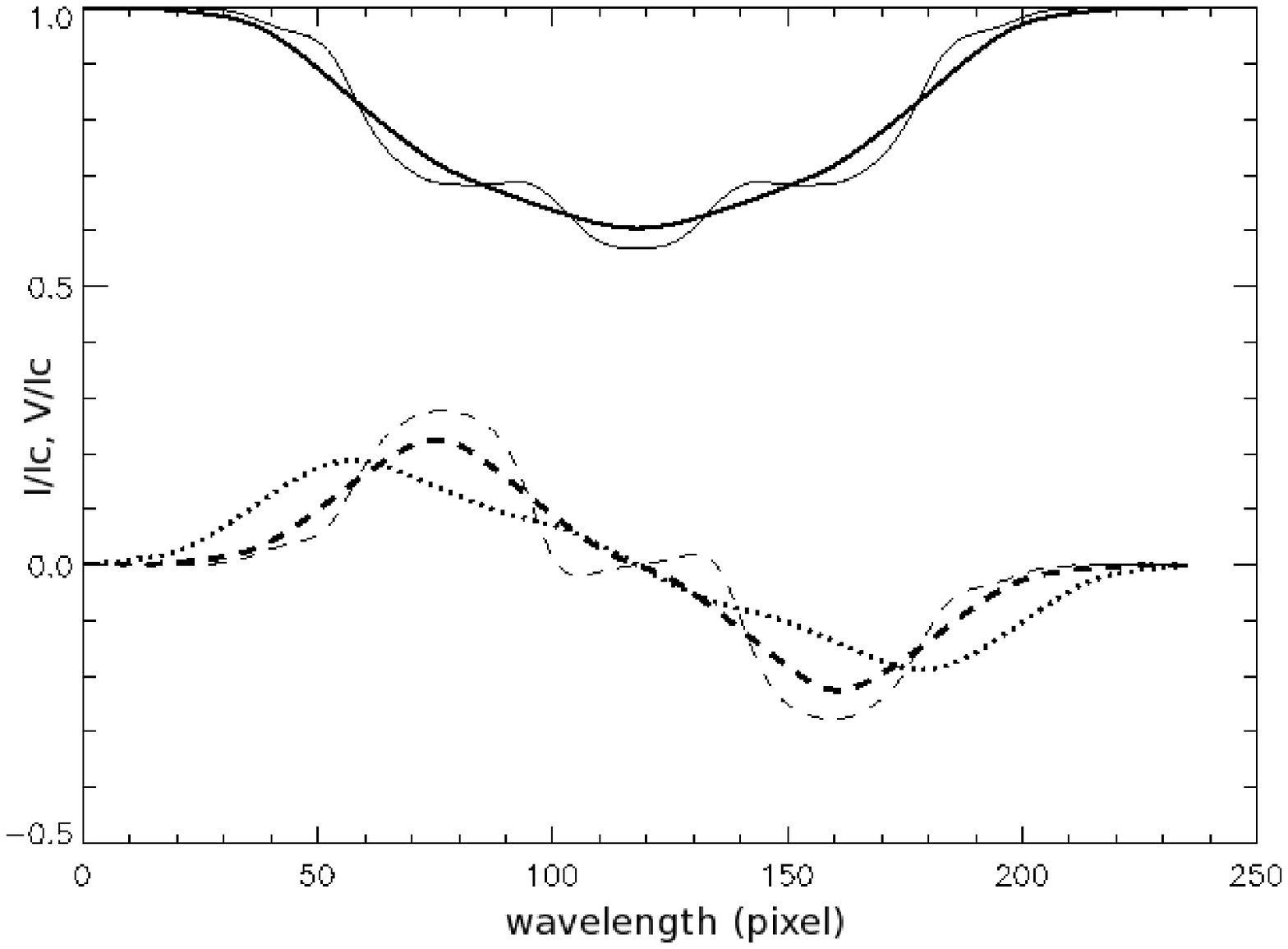}
\caption{Top panels: spectral synthesis of the 9 selected lines of the multiplet 816 of Fe\,{\sc i} 
(see Tab. \ref{table1}; line 1 to 9 from left to right) having a spectral resolution of 3$\times$10$^6$ 
and a sampling of 100\,m s$^{-1}$ (corresponding to steps of about $2.1$\,m\AA\ in wavelength). 
From left to right the magnetic field strength is $B$ = 500, 1500 and 2500 G. The inclination of the 
magnetic field vector with respect to the line of sight is the same 
for all the columns, having a value of $50^{\circ}$. Bottom panels: The MZS resulting from the addition of the
nine spectral lines. In all panels, the upper curves in solid line show Stokes $I$ profiles while 
the lower curves in dashed line represent Stokes $V$. The thin lines are the MZS having a resolution of 
3$\times$10$^6$. The thick lines represent the MZS smoothed to reduce 
the spectral resolution to 75000. Additionally, the bottom panels also show the derivative of the smoothed low 
resolution intensity profile in dotted line.}
\label{fig:mzs_fields}
\end{figure*}

Assuming wavelength-independent Gaussian noise in the observations
characterised by a variance $\sigma^2$, the observed Stokes $V(x_j-X_i)$ profiles can be
considered to be random variables characterised by a normal
distribution with a mean value of $\mu(x_j)$ and variance $\sigma^2$. In our case,
$\mu(x_j)$ is the hidden Zeeman signature at Doppler coordinate $x_j$ that we want to recover.
According to the results presented in Fig. \ref{fig:weights}, 
the MZS of Eqs. (\ref{eq:LA}) and (\ref{eq:LSD}) can be approximated
to be the addition of $N$ products of two Gaussian random variables. 
It is known that the probability distribution of the product
of two Gaussian random variables $X$ and $Y$ is, in general, very complex to calculate.
However, it is possible to obtain the mean and variances of such product
through the calculation of the moment generating function (Craig 1936):
\begin{eqnarray}
E(XY) &=& \mu_X \mu_Y \nonumber \\
V(XY) &=& \mu_X^2 \sigma_Y^2 + \mu_Y^2 \sigma_X^2 + \sigma_X^2 \sigma_Y^2,
\label{eq:mean_variance}
\end{eqnarray}
where the $\mu_{\{X,Y\}}$ represent the means of the two variables while the $\sigma_{\{X,Y\}}^2$ are
their corresponding variances.
In our case, $X$ corresponds to the normalised weights, so that $\mu_X=1/N$ and $\sigma_X^2=\sigma_w^2$, with
$\sigma_w$ the standard deviation of the weights. Likewise, $Y$ corresponds
to the observed Stokes $V$ profiles, so that $\mu_Y=\mu(x_j)$ and $\sigma_Y^2=\sigma^2$. For calculating the MZSs, we have
to add many such products. At this point, following the central limit theorem, the resulting 
probability distribution function can be calculated as the convolution of $N$ Gaussian distributions
whose mean and variance are given by Eq. (\ref{eq:mean_variance}). The resulting mean is just $N$ times
the mean of each Gaussian while the resulting variance is $N$ times the variance of each individual
Gaussian. Therefore:
\begin{equation}
Z_\mathrm{LSD}(x_j) \approx Z_\mathrm{LA}(x_j) \approx Z_\mathrm{SLA}(x_j) \approx \mu(x_j).
\label{eq:ZLSD_final}
\end{equation}
with variance
\begin{equation}
\sigma_Z^2(x_j) = N \sigma^2 \left( \frac{1}{N^2} + \sigma_\omega^2 \right) + N\mu(x_j)^2 \sigma_\omega^2.
\label{eq:ZLSD_variance}
\end{equation}
The signal-to-noise ratio increases roughly linearly
with the number of added lines but reaches a maximum value at:
\begin{equation}
N_\mathrm{max} = \frac{\sigma}{\sigma_\omega \sqrt{\mu(x_j)^2 + \sigma^2}} = \frac{1}{\sigma_w \sqrt{1+(S/N)_\mathrm{real}^2}},
\label{eq:Nmax}
\end{equation}
where $(S/N)_\mathrm{real}=\mu(x_j) / \sigma$ is the real signal-to-noise ratio, i.e. the ratio between the
amplitude of the hidden Zeeman signature and the noise. Equation (\ref{eq:Nmax}) demonstrates
that $N_\mathrm{max}$ represents the maximum number of lines one should add in order to have a net increase
in the signal-to-noise ratio. If more lines are added, the signal-to-noise ratio starts to
degrade because of the intrinsic dispersion in the weights. The larger the ratio
$\mu(x_j) / \sigma$, the smaller the number of lines one should add until reaching
the regime of $S/N$ degradation. Likewise, the larger the dispersion of the weights, the smaller
the number of lines one should add.

The previous results suggest the surprising possibility of \emph{using purely random weights in
the application of line addition or LSD}. The only requisite is that the weights are 
positive definite with a fairly symmetric distribution peaking close to $1/N$ with
a sufficiently small dispersion not to dominate in Eq. (\ref{eq:ZLSD_variance}). Such 
a distribution should lead to results similar to
those presented in Eqs. (\ref{eq:ZLSD_final}) and (\ref{eq:ZLSD_variance}). Also surprising
is the fact that the analysis we have carried out suggests that the best solution is just
to add the lines without any weight. In this case, no saturation due to the presence of
dispersion in the weights appears. This is a consequence of the fact that, since the 
observations are assumed to be described by the Gaussian distribution $N(\mu(x_j),\sigma^2)$, 
the maximum-likelihood estimation of its mean is exactly given by Eq. (\ref{eq:simplified_LA}).

Summarising, one can consider that LSD and complex line addition schemes work because
they give a response similar to that of the simplest line addition method. The fundamental
reason is that spectral lines constitute a \emph{well-behaved statistical sample}. The line-to-line
differences are small enough (for instance, the dispersion in the value of the LSD or LA
weights) so that the plain addition of lines makes the signal appear
as soon as one adds a sufficient number of lines. Although the assumptions on which
LSD or any method are not really fulfilled, the addition of lines dilutes any possible
difference in their behaviour.

\section{The magnetic sensitivity of the MZS}
The MZS are very efficient detectors of the magnetic activity of the star by means of the 
Zeeman effect. However, since our main interest is to understand the magnetism of the stars, 
we need to infer the magnetic field vector of the star from the information 
contained in the polarised spectra or from the MZS. In cools stars the signal to noise ratio 
per individual spectral line is usually so poor that it is impossible to make 
reliable measurements of the magnetic field. Consequently, we are obliged to use the MZS as indicators 
of the magnetic activity in the stars. At this point a question arises: does the MZS contain 
the information of the magnetism of the star? It is evident that, in the case of the line 
addition, the MZS is not a physical spectral line. Then, does it behave like a spectral line 
and contain any information about the magnetic field?. This section will be devoted to study 
to what extent it is possible to use the MZS to reliably infer the magnetic field in stars.


We simulate the observed spectral lines from a star assuming a Milne-Eddington 
atmosphere. Moreover, we suppose that there is only one magnetic field vector in the 
stellar surface that is responsible for the Zeeman signatures in the spectra. 
Although more sophisticated scenarios and solutions to the polarised radiative transfer 
equation exist, we are not interested in a realistic simulation of the stellar spectra 
but in showing that, once the information of the spectra is encoded in the MZS, we are 
capable of retrieving it. Using the M-E model facilitates the analysis and does not 
require heavy calculations. We use the code {\it Diagonal}
as described in L\'opez Ariste and Semel (1999) with only one layer. This code 
can arbitrarily cope with a large number of layers if needed. For the sample M-E model atmosphere 
we choose the value of the gradient of the source function to be
$\beta = 10$ and the ratio between the line and continuum absorption, $\eta$, 
to be equal to twice the relative strength of the 
components of the multiplet as given by Allen (2000). Doppler broadening 
is set to $21$\,m\AA\ as in sunspots, and the reduced damping is taken to be 0.01.\footnote{Saturation 
has always been an important issue in solar and 
stellar magnetic field measurements. Here it is included in the
procedure and we have to use the same procedure when we get to the 
inversion.}

We synthesise the Stokes vector of only a few spectral lines with a spectral 
resolution close to 3$\times$10$^{6}$. We select the multiplet 816 of
Fe\,{\sc i} (listed in Tab. \ref{table1}) to ensure the validity of LS coupling 
and the easy determination of Zeeman patterns, relative line strengths, etc. We then add 
up the Stokes parameters of all the spectral lines using unit weights [therefore, using
the simplified line addition of Eq. (\ref{eq:simplified_LA})] and subsequently reduce the spectral 
resolution to stellar conditions, say 75000, obtaining the MZS. 

\begin{table}[!t]
 \begin{tabular}{ccccccr}        \hline
&Line  & $\lambda$  & upper &   lower  & relative  & $\eta$\\
& No.  & \AA        & level &   level  & strength &       \\
 \hline
& 1  & 6400.010     & $^{5}D_4$   &    $^{5}P_3$   &    27     & 54   \\
& 2  & 6411.658     & $^{5}D_3$   &    $^{5}P_2$   &    14     & 28   \\
& 3  & 6408.031     & $^{5}D_2$   &    $^{5}P_1$   &    5.25   & 10.5 \\
& 4  & 6246.334     & $^{5}D_3$   &    $^{5}P_3$   &    7      & 14   \\
& 5  & 6301.515     & $^{5}D_3$   &    $^{5}P_2$   &    8.75   & 17.5 \\
& 6  & 6336.835     & $^{5}D_1$   &    $^{5}P_1$   &    6.75   & 13.5 \\
& 7  & 6141.734     & $^{5}D_2$   &    $^{5}P_3$   &    1.     & 2    \\
& 8  & 6232.661     & $^{5}D_1$   &    $^{5}P_2$   &    2.25   & 4.5  \\
& 9  & 6302.507     & $^{5}D_0$   &    $^{5}P_1$   &    3.     & 6    \\
\hline
\end{tabular}
\caption{The list of spectral lines of multiplet 816 of neutral iron.
The relative strengths correspond to LS coupling as given by Allen (2000).
The ratio $\eta$ of line to continuum absorption is obtained 
from the relative strengths multiplied by a factor of 2; the latter
is chosen arbitrarily.}
\label{table1}
\end{table}

\begin{figure*}[!t]
\center
\includegraphics[width=0.4\textwidth]{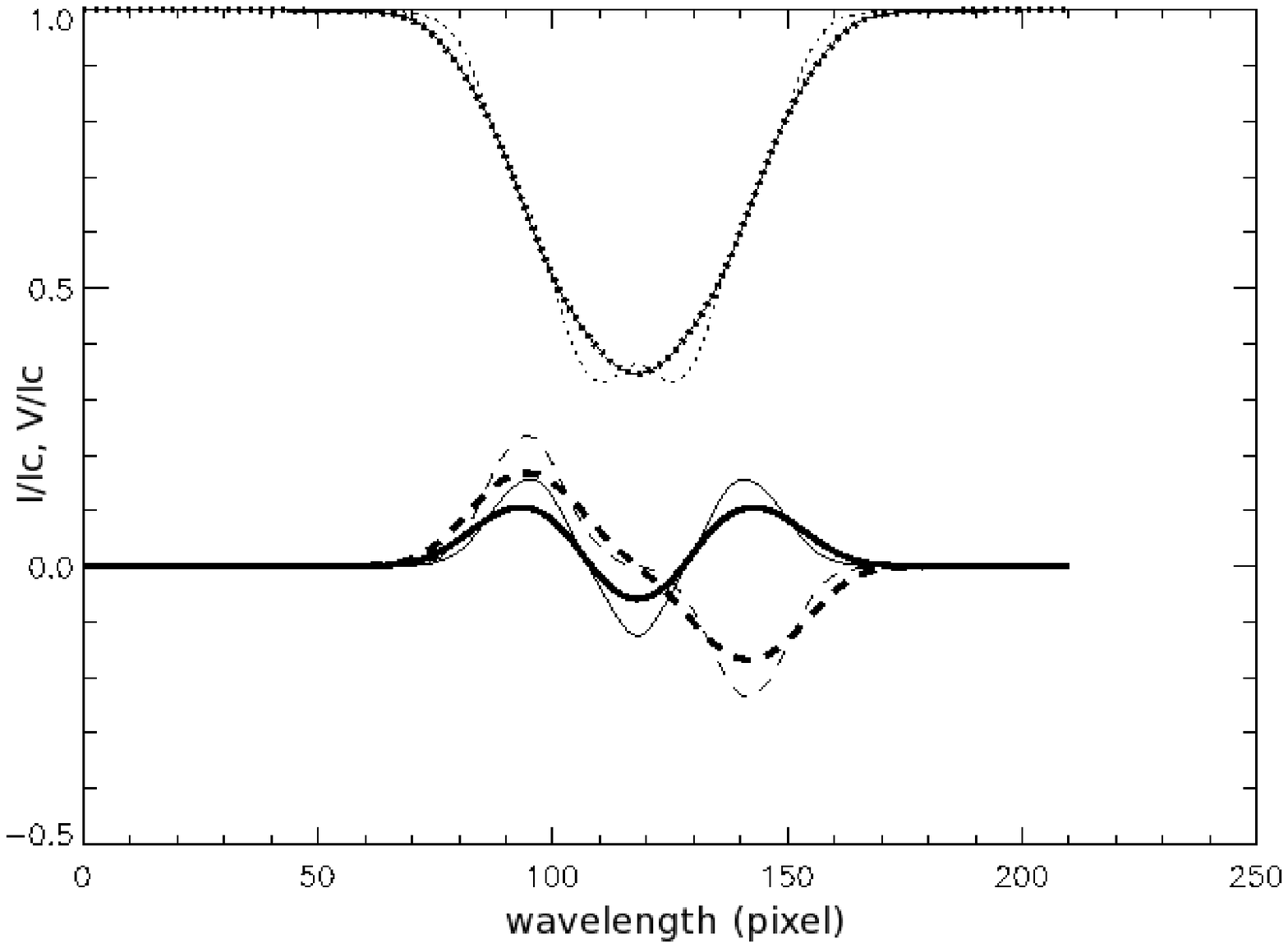}
\includegraphics[width=0.4\textwidth]{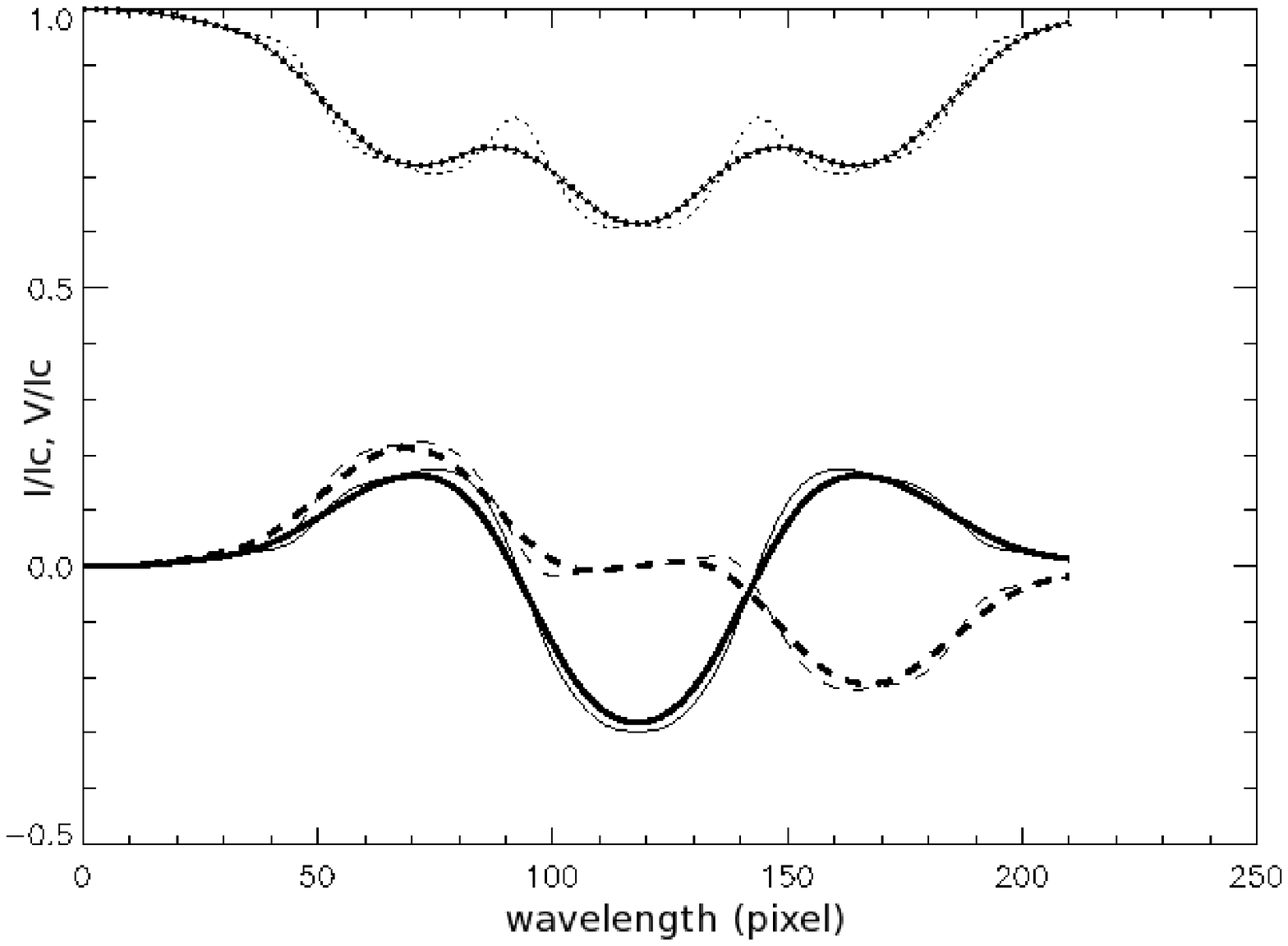}\\
\includegraphics[width=0.4\textwidth]{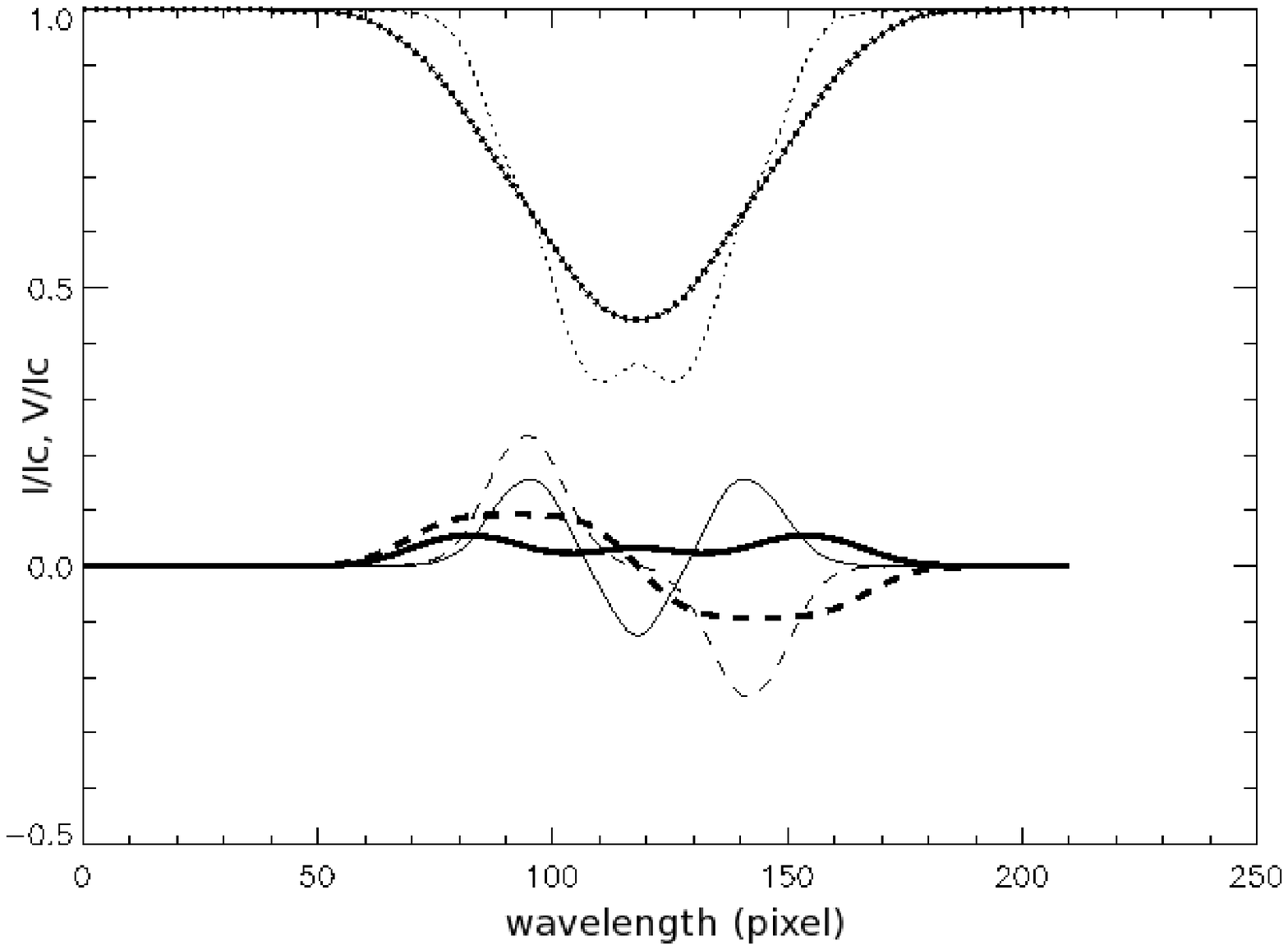}
\includegraphics[width=0.4\textwidth]{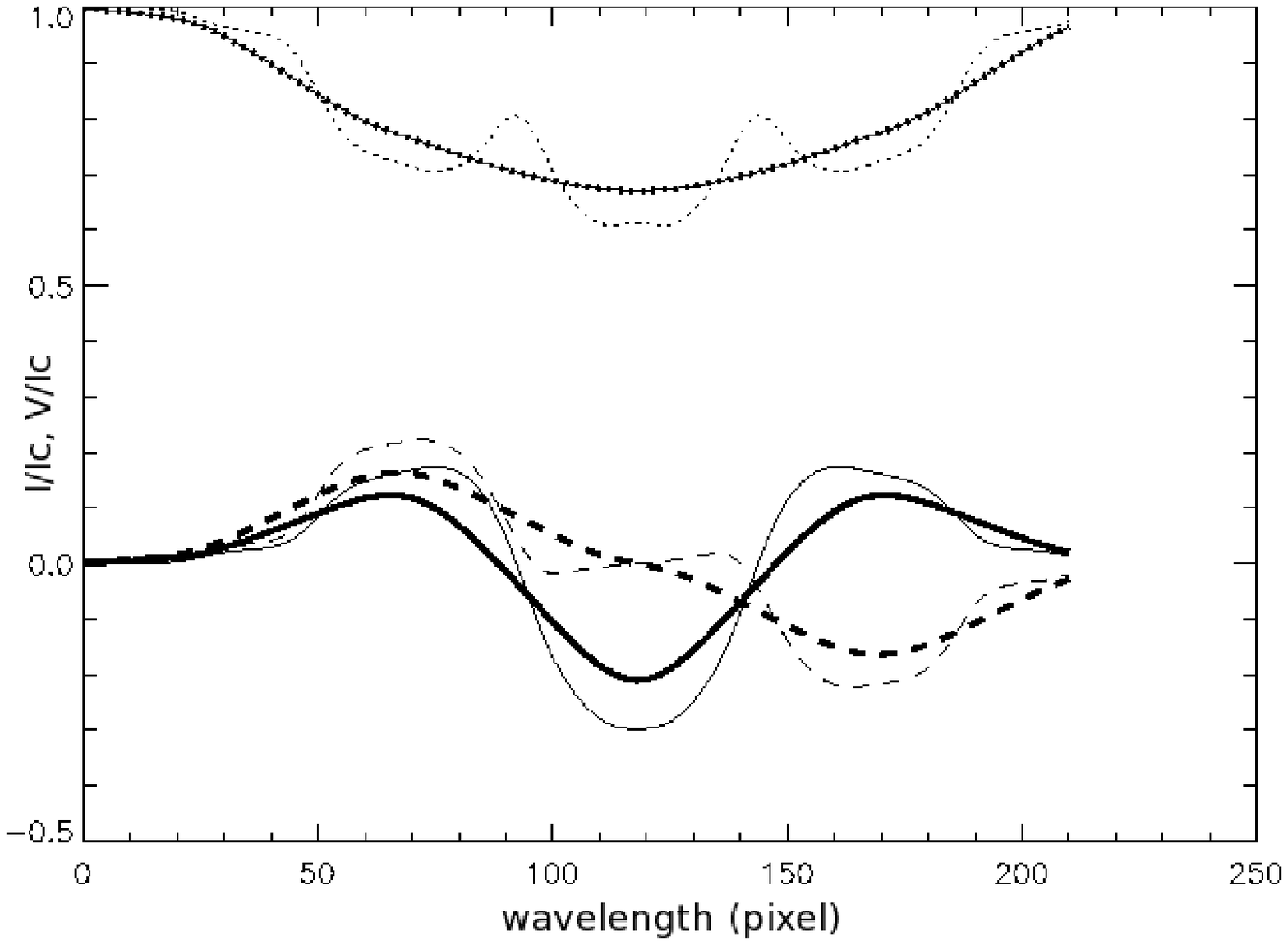}
\caption{The MZS for a magnetic field strength of 1000 G (left panels) and 3000 G (right panels). Inside each panel,
the intensity profile are shown at the top, while the polarisation profiles are shown at the bottom, with
Stokes $V$ shown in dashed lines and Stokes $Q$ in solid lines. High spectral resolution profiles are
shown in thin lines, while the smeared low spectral resolution profiles are indicated with thick lines.}
\label{fig:figure_stokesQ}
\end{figure*}

\subsection{MZS in circular polarisation}

Figure \ref{fig:mzs_fields} shows the synthetic Stokes $I$ and $V$ together with the MZS for the selected Fe\,{\sc i} lines 
for three different values of the magnetic field strength. We keep a high spectral resolution of 100\,m/s, 
i.e. a resolution of 3\,millions, and then study the effect on the MZS of smoothing, bringing the resolution 
down to 75000, which corresponds to sampling with a step of about 4\,km s$^{-1}$.

With very high spectral resolution both Stokes $I$ and Stokes $V$ profiles 
change morphology and details clearly depending on the field 
strength. In the case of 500 G we do not detect a clear magnetic splitting in the Stokes $I$ profiles. Also, the 
Stokes $V$ profiles are all very similar, the only difference being a scale factor. This is a 
consequence of the fact that for 500 G, all spectral lines are in the weak field regime of the Zeeman effect,
where the magnetic splitting is smaller than the Doppler width of the line. However, when the 
magnetic field increases, the Zeeman splitting becomes evident in the intensity profiles and the circular 
polarisation profiles abandon the shape-similarity regime. In fact, the anomalous dispersion terms are 
now important and show the typical reversal of polarities in the centre of the Stokes $V$ profile. 
This has no counterpart in the weak-field approximation since anomalous dispersion is a high-order
effect. Although the spectral lines are now out of the weak field regime, they have not still reached 
the strong field regime (where the Zeeman splitting is very large as compared to the 
Doppler width of the line). 

Considering the MZS, in the weak field case (500 G), we see that, even in the high spectral resolution case, 
the shapes of the intensity and circular polarisation MZS also resemble the individual Stokes $I$ and 
Stokes $V$ profiles, respectively. When lowering the spectral 
resolution to the typical stellar one, the profiles are also similar in shape but the amplitude has decreased 
and the profiles are wider. Note that the MZS in circular polarisation matches almost perfectly with the 
derivative of the intensity profile. The reason is simple to understand. Since all the lines
are in the weak-field regime and the derivative is a linear operation, the addition of derivatives
equals the derivative of the addition of lines. This confirms that in this case the weak field regime 
is a good approximation. 

When we leave the weak field regime, the MZS at high spectral resolution shows some distortions 
that are due to the different shapes of all the individual Stokes $V$ profiles (every line departs
from the weak-field regime at different values of the magnetic field strength). When we decrease the 
spectral resolution to the stellar case, in the case of 1500 G we see that the MZS for circular polarisation 
is close to the derivative of the intensity MZS profile. This means that, even if the individual spectral 
lines are not in the weak field regime, the MZS for circular polarisation can be in this regime when 
lowering the spectral resolution. The reason behind this behaviour is that the width of the lines
are increased when lowering the spectral resolution, thus extending the validity of
the weak-field regime. Consequently, the weak field regime applies for larger magnetic fields in 
the circular polarisation MZS than in the individual spectral lines (typically up to 600-700 G in the 
visible part of the spectrum). However, this does not occur in the 2500 G case. The low resolution circular 
polarisation MZS already differs substantially from the derivative of the intensity MZS.

The MZS constructed with line addition are not simple spectral lines in the sense that they have 
no associated wavelength and are not characterised by specific atomic parameters 
such as the excitation potential. Moreover, in the more general case they 
do not even correspond to a particular chemical element, and no specific 
spectroscopic term is attached to them. However, the Zeeman signatures do not disappear 
in the MZS. We have seen that they are sensitive to the magnetic field, having also a weak field 
regime where the amplitude of the circular polarisation MZS scales with the derivative of the intensity MZS.
For this reason, they can be used to extract information about the magnetic field, 
as we show in \S\ref{sec:magnetic_info}.

\subsection{MZS in linear polarisation}
As a rule of thumb, the Zeeman linear
polarisation is at least one order of magnitude lower than the circular one (e.g. Carroll et al. 2007).
Now, if our line of sight (henceforth LOS) is not a preferred direction 
for the stellar magnetic field vector and therefore its three components  
(the longitudinal one and the two perpendicular to the LOS) have all equal 
probabilities, on average the transverse component is $\sqrt 2$
times stronger than the LOS component. 
From solar physics we know that kG fields are not only found in sunspots. 
However, can we assume that kG fields are common also in solar type 
stars? If yes, why does the linear polarisation escape observation? In 
the following, we try to answer this question and we also suggest a remedy.

Let us denote $\psi$ the angle between the magnetic field vector and the LOS and take it between $0^{\circ}$ 
and $90^{\circ}$, say positive LOS component. If all directions have the
same probability, the average inclination angle is $\overline{\psi}=1$\,rad $\approx 57.3^\circ$. Likewise, the 
median angle is  $\psi_\mathrm{Median} = 60^{\circ}$. 
In other words, there are as many orientations 
between $0^{\circ}$ and $60^{\circ}$ as between $60^{\circ}$ and $90^{\circ}$. 
For the case of negative LOS longitudinal fields,
$90^{\circ}<{\psi}< 180^{\circ}$, the results for  $\overline{\psi}$ and for  
$\psi_{Median}$ are deduced similarly. In conclusion, $\psi$ is likely to 
be nearly $60^{\circ}$. 

Figure \ref{fig:figure_stokesQ} shows the MZS for the Stokes $I$, $Q$ and $V$ parameters computed 
with a magnetic field strength of 1\,kG and 3\,kG and an inclination $\psi = 60^{\circ}$. 
Note that, for high spectral resolutions, the MZS for Stokes $Q$ is approximately 0.7 times the MZS for Stokes $V$. 
Stokes $Q$ changes with wavelength twice as fast as Stokes $V$, so that  
for current high spectral resolutions, say 75000, Stokes $Q$ shrinks much faster than Stokes $V$ in the convolution
process. Moreover, Stokes $Q$ changes sign when the azimuth changes by $90^{\circ}$, while Stokes $V$ 
changes sign when $\psi$ changes by $180^{\circ}$. Both effects can be 
reduced by increasing the spectral resolution, to values of the order of $120000$. 
This means that observing at high spectral resolutions, Stokes $Q$ is preserved considerably 
and, more interestingly, the spatial resolution of the stellar surface is improved as well.

\section{Recovering the magnetic field from the MZS}
\label{sec:magnetic_info}
Our purpose is to assess to what extent the new MZS still contains
all the information on the magnetic field that was present in the 
individual spectral lines in our sample. To this end, we examine the invariants 
that we have already found in solar magnetism, like the displacement of the centre 
of gravity (CoG) that yields a good approximation to the longitudinal component of the magnetic field. 
It is important to have in mind that the centre of gravity is a linear operator and commutes with the
algebra of line addition and with smoothing.

The centre of gravity method applies only to circular polarisation; it is well studied 
in solar magnetism and described in a number of papers of which at
this point we mention only Rees et al. (1979) (and references therein). In the limiting case
of optically thin layers in local thermodynamical equilibrium, the centre of gravity shift of the 
line profiles observed in circular polarisation is proportional to
the longitudinal component of the magnetic field. This still holds 
true, even for line formation in optically thick layers, in the 
limiting case of weak magnetic fields. Usually, this simple method
is still a fair approximation in the more general case, but there 
are a few exceptions -- see Semel (1967) for theoretical 
demonstrations, experimental tests and some exceptions. For further 
discussion of the CoG method applied to stars see also Stift (1986) and Leone \& Catanzaro (2004).

\begin{figure}[!t]
\resizebox{9cm}{!}{\includegraphics{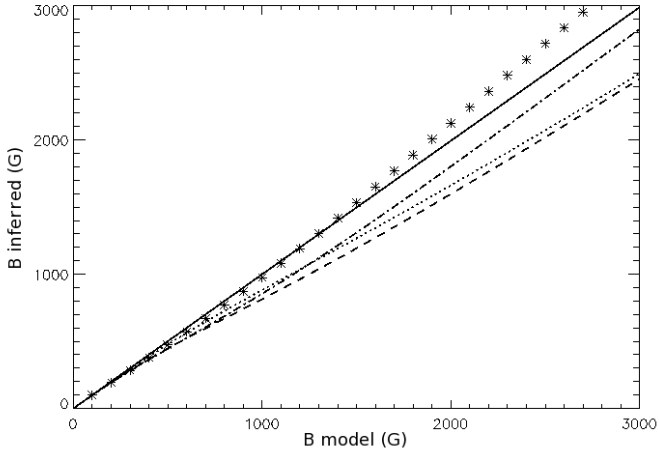}}
\caption{\textit{Top:} A test of the centre of gravity method. The CoG
shifts are calculated over a database of 30 values of magnetic fields in 
steps of 100\,Gauss from 0 to 3\,kG. The shifts are divided by the cosine
of the inclination angle and by the adopted effective Land\'e factor to compare with 
the amplitude of the magnetic field (abscissa). The curves are shown 
for the following values of $\psi$, the inclination angles in degrees:
$0$ (solid line), 20 (points),  40 (dashed), 60 (dashed - points)
and 80 (stars). The value of $\bar g$ for the pseudo-line was chosen to be
1.59, For $\psi = 0$ the CoG shifts correspond  exactly to the line
of sight component of the magnetic field.}
\label{fig:cog}
\end{figure}

We apply the centre of gravity method to the MZS of Stokes $V$ 
(the ones showed in Fig. \ref{fig:mzs_fields}). We compute the MZS for Stokes $V$ assuming 
a magnetic field strength going from 0 to 3\,kG having different inclinations 
${\psi}=0, 20, 40, 60$ and 80$^\circ$. Then, we infer the magnetic field 
following the centre of gravity method. Figure \ref{fig:cog} shows the plot 
of the deduced magnetic field versus the theoretical one. The figure shows a clear linear trend 
showing that the MZS contains the information of the magnetic field that was present in the 
individual spectral lines. Since the centre of gravity is applicable both to the individual spectral lines and
to the MZS of Stokes $V$, we conclude that the MZS encodes the information of the magnetism of the 
star in a similar way that the individual spectral lines do. In other words, the MZS constructed with the 
line addition technique behaves as a spectral line although not any specific atomic element is associated to it. However, 
it can be characterised by an effective land\'e factor and the extraction of the magnetic information 
is possible. Figure \ref{fig:cog} also shows an error close to 10\% of the inferred value which is intrinsic to the 
centre of gravity method. Our aim here is to show that the magnetic information is present in the MZS and not to 
do precise measurements of the magnetic field strength. In order to do these exact measurements we must use 
more sophisticated techniques. To have reliable values of the magnetic field, it is fundamental that \emph{the 
inversion procedure uses exactly the same operator {\sf O} that generates the MZS
for the comparison with the observational data}. This 
topic will be discussed in a forthcoming paper.

\section{Conclusions and discussion}


We have used the line addition technique to show how to extract a Zeeman 
signature for any of the Stokes parameters. These technique, apart from being 
applicable to any state of polarisation, is also model independent. In other words, 
we do not need to do any approximation nor any assumption on the magnetic field or the 
stellar atmosphere to extract the Zeeman signature from the spectra. The line addition, 
as simple as the brute force addition of many spectral lines is a powerful 
tool to detect the magnetic field in cool stars, where the Zeeman signature per spectral line 
is very poor. 

Moreover, the MZS created by the line addition technique also contains the 
magnetic information of the star. Therefore, it is possible to infer the 
magnetic (and the thermodynamical) properties of the stellar atmosphere using the 
appropriates inversion procedures. We have shown that the magnetic field strength 
can be inferred from the MZS of Stokes $V$ by using a procedure as simple as the centre of gravity. 
However, we must point out that, in order to have more reliable measurements of the 
magnetic field vector, we need to apply more sophisticated inversion procedures. Such
procedures should ideally take into account in the process the same operator used to 
create the MZS when comparing with the observations.

Linear polarisation is often neglected in stellar polarimetry under 
the assumption that it will present amplitudes too small to 
be observed. In this paper, we have demonstrated that, at 
appropriate spectral resolutions, the linear polarisation due to the Zeeman 
effect produces sensible signatures useful for diagnostics of the stellar magnetic fields.
With a resolution above 10$^5$, we can probably detect significant 
signals in linear polarisation in the spectra of solar type stars. 
Such a resolution has been achieved in spectroscopic tests with the SemPol instrument working 
together with UCLES at the AAT with the 79\,G/mm grating\footnote{See the discussion by by L\'opez Ariste \& Semel (1999)
on the spectral resolution of UCLES in \texttt{http://www.aao.gov.au/local/www/UCLES/cookbook/report2.html}.}.
The detection of linear polarisation in the spectrum 
of a solar type star has been performed in 2004 (see Semel et al, 2006). The
linear signal observed was indeed four time less than the circular one, 
but still significant. There is definitely an interest to proceed in 
this direction.

We have also demonstrated why line addition and least-squares deconvolution give
very similar MZS. The fundamental reason is that the weights used in both cases
have well-defined statistical properties: their probability distribution is
positive definite, centred at the inverse of the number of lines and not too
broad. These properties induce that weighted line addition and LSD converge to
the simplest possible line addition which, since noise is assumed to be of Gaussian type,
turns out to be the maximum-likelihood estimation of the MZS. This opens the
curious possibility of applying LSD with random weights provided they fulfil
the mentioned properties.

In this paper we have presented a very simple technique to build the MZS. However, more powerful 
techniques are being developed in the frame of PCA. The main 
idea is to build a more sophisticated operator than the simple Dirac functions used for the 
line addition. The operator using PCA will be related to the eigenvectors of a data base containing 
theoretical stellar spectra synthesised in several magnetic and thermodynamic configurations. 
Ramirez V\'elez et al. (2009, submitted) explains this technique in detail and apply recent solar inversion procedures
to the synthesised MZS using PCA. Other applications of PCA to obtain the MZS
has been presented by Mart\'{\i}nez Gonz\'alez et al. (2008) and by
Carroll et al. (2007). Application of PCA has been proven very effective 
in solar magnetometry and, analogously to the solar case, a database will be 
created and 
the PCA eigenvectors derived, applying the PCA-ZDI detection to just one 
magnetic point on the star. In later stages of the work, 
we shall discuss how to apply our 
approach to some few points on the stellar disc well separated by 
significant Doppler effects and, finally, we shall treat 
the more realistic case of continuous distributions of fields over the stellar surface. 
Here, the orthogonality of the eigenvectors is not conserved for adjacent 
stellar points (that is for small differences in the Doppler shifts).
 
\acknowledgements
Dr. David Rees has introduced PCA methods to the field of solar magnetometry 
which was the starting point for this method in general magnetometry 
including solar type stars.
Here,  one of us (M. Semel) wishes to express his gratitude
to Dr. David Rees for lectures on PCA given during his visit to Meudon 
Observatory in the year 2002.
We thank the referee, Prof. J. Landstreet, for many valuable comments that helped
improve the paper.
Our thanks go to Prof. S. Cuperman for reading the paper and for very 
precious comments and corrections. 
MJS acknowledges support by the Austrian Science Fund (FWF), 
project P16003-N05 ``Radiation driven diffusion in magnetic stellar
atmospheres'' and through a Visiting Professorship at the Observatoire
de Paris-Meudon and Universit{\'e} Paris 7 (LUTH).
MJMG and AAR acknowledges finantial support by the Spanish Ministry of Education and 
Science through project AYA2007-63881. AAR also acknowledges the European Commission through 
the SOLAIRE network (MTRN-CT-2006-035484).

\section {References} 
Allen, C. W. 2000, Astrophysical Quantities (New York: Springer Verlag and AIP Press)\\
Carroll, T. A., Kopf, M., Ilyn L., \& Strassmmeier K. G. 2007, Astron. Nachr., 88, 789. \\
Brown, S. F., Donati, J.-F., Rees, D. E., \& Semel, M. 1991, A\&A 250, 463\\
Craig, C. C. 1936, Ann. Math. Statist., 7, 1 \\
Donati, J.-F., Brown, S. F. 1997, A\&A 326, 1135\\
Donati, J.-F., Semel, M., Carter, B. D., Rees, D. E., \& Collier Cameron, A. 1997, MNRAS, 291,658 \\
Kurucz, R. L. 1993, Phys. Scr. T, 47, 110\\
Landi degl'Innocenti, \& Landolfi, M. 2004, Polarization in Spectral Lines (Dordrecht: Kluwer Academic Publishers)
Leone F., \& Catanzaro, G. 2004, A\&A, 425, 271\\
L\'opez Ariste, A., \& Semel M. 1999, A\&AS, 139, 417\\
L\'opez Ariste, A., \& Semel M. 1999, WEB page :  \\
Mart\' inez Gonz\'alez, M. J., Asensio Ramos, A., Carroll, T. A., Kopf, M., Ram\' irez V\'elez, J., \& Semel, M. 2008, A\&A, 486, 637\\
Rees, D. E. and Semel, M. D. 1979, A\&A, 74, 1 \\
Rees, D. E., L\'opez Ariste A., Thatcher J., \& Semel M. 2000, A\&A, 355, 759\\
Sears F. H. 1913, ApJ, 38, 99\\
Semel, M. 1967, A\&A, 30, 257\\
Semel, M. 1980, A\&A, 91, 369\\
Semel, M. 1987, A\&A, 178, 257\\
Semel, M. 1989, A\&A, 225, 456\\
Semel, M. 1995, in ASP Conf. Ser., Vol. 71, Tridimensional Optical Spectroscopic Methods in Astrophysics, ed.
G.~{Comte} \& M. {Marcelin}, 340\\
Semel M. and Li J. 1996, Sol. Phys., 164, 417\\
Semel M., Rees D.E., Ram\'{\i}irez V\'elez, J. C., Stift, M. J., \& Leone F., in ASP Conf. Ser., Vol. 358, Solar Polarization 4, ed.
  R.~{Casini} \& B.~W. {Lites}, 355\\
Stift, M. J. 1986, MNRAS, 221,499\\
Stift, M. J. 2000, COSSAM: Codice per la Sintesi Spettrale nelle Atmosfere Magnetiche, A Peculiar Newsletter, 33, 27\\

\end{document}